\def\etcl{$\kappa$-(BE\-DT\--TTF)$_2$\-Cu\-[N\-(CN)$_{2}$]Cl}
\def\etcn{$\kappa$-(BE\-DT\--TTF)$_2$\-Cu$_2$(CN)$_{3}$}
\def\cm{cm$^{-1}$}
\documentclass[prb,twocolumn,showpacs,amsmath,amssymb]{revtex4}
\usepackage[dvips]{graphicx}
\begin{document} 
\title{Power-law dependence of the optical conductivity
observed in the \\
quantum spin-liquid compound $\kappa$-(BE\-DT\--TTF)$_2$\-Cu$_2$(CN)$_{3}$}
\author{Sebastian Els\"asser}
\affiliation{1.~Physikalisches Institut, Universit\"{a}t
Stuttgart, Pfaffenwaldring 57, D-70550 Stuttgart Germany}
\author{Dan Wu}
\affiliation{1.~Physikalisches Institut, Universit\"{a}t
Stuttgart, Pfaffenwaldring 57, D-70550 Stuttgart Germany}
\author{Martin Dressel}
\email{dressel@pi1.physik.uni-stuttgart.de}
\homepage{http://www.pi1.physik.uni-stuttgart.de}
\affiliation{1.~Physikalisches Institut, Universit\"{a}t
Stuttgart, Pfaffenwaldring 57, D-70550 Stuttgart Germany}
\author{John A. Schlueter}
\affiliation{Material Science Division, Argonne National Laboratory,
Argonne, Illinois 60439-4831, U.S.A.}
\date{\today}
\begin{abstract}
The Mott-insulator $\kappa$-(BEDT-TTF)$_2$Cu$_2$(CN)$_3$ is the
prime candidate of a quantum spin liquid with puzzling magnetic
properties. Our THz and infrared investigations reveal that also
the charge dynamics does not follow the expectations for a Mott
insulator. The frequency-dependent conductivity exhibits a
power-law behavior $\sigma_1(\omega)\propto \omega^n$ that grows
stronger as the temperature decreases and extends all the way
through the far-infrared. With $n\approx 0.8$ to 1.5 we obtain a
significantly smaller exponent than predicted by Ng and Lee [Phys.
Rev. Lett. {\bf 99}, 156402 (2007)]. We suggest fluctuations
becomes important in the spin-liquid state and couple to the
electrodynamic properties differently compared to the
antiferromagnetic Mott insulator
$\kappa$-(BE\-DT\--TTF)$_2$\-Cu\-[N\-(CN)$_{2}$]Cl. We discuss the
various possibilities of how charge fluctuations are influenced by
the presence or absence of magnetic order.
\end{abstract}


\maketitle
%
%

\section{Introduction}
\begin{figure}
 \centering
 \includegraphics[width=1\columnwidth]{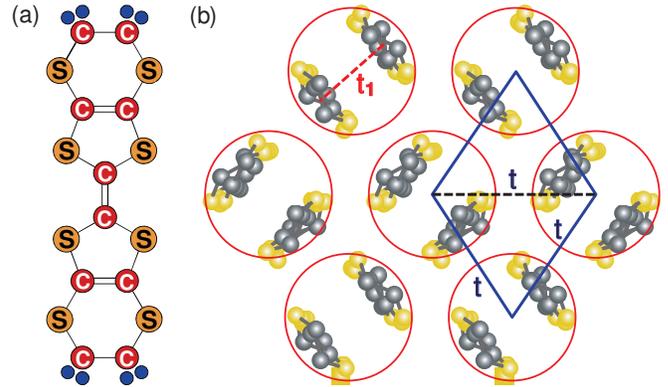}
\caption{\label{fig:1} (Color online) (a) Sketch of the
BEDT-TTF molecule. (b) For $\kappa$-(BEDT-TTF)$_2X$ the molecules
are arranged in dimers, which constitute an anisotropic triangular
lattice within the conduction layer. The inter-dimer transfer
integrals are labeled by $t$ and $t^{\prime}$ and can be
calculated by tight-binding studies of molecular orbitals or
ab-initio calculations.\cite{Nakamura09,Kandpal09,Jeschke12} With
the intra-dimer transfer integral $t_1\approx 0.2$~eV
\cite{Oshima88,Komatsu96} and the onsite Coulomb repulsion
$U\approx 2t_1$,\cite{McKenzie98} one obtains at ambient
conditions $U/t=5.5$ with the ratio of the two inter-dimer
transfer integrals $t^{\prime}/t \approx 0.44$  in the case of the
Mott insulator \etcl. For the spin-liquid compound \etcn, the
effective Hubbard $U$ is larger ($U/t=7.3$) and most important the
transfer integrals $t^{\prime}/t=0.83$ are very close to equality.
}
\end{figure}

Among the two-dimensional organic charge-transfer compounds, the
$\kappa$-(BEDT-TTF)$_2X$ family [where BEDT-TTF stands for
bis-(ethyl\-ene\-di\-thio)\-te\-tra\-thia\-ful\-va\-lene]
is of
particular interest, because the constituting cationic dimers are
arranged in an anisotropic triangular lattice  (Fig.~\ref{fig:1})
with a delicate interplay between electronic correlations (given
by the on-site Coulomb repulsion $U$), the effects of low
dimensionality and spin frustration.\cite{Kanoda11,Powell11}
While \etcl\ is a Mott insulator with antiferromagnetic order at
low temperatures ($T_N\approx 25$~K),\cite{Welp92,Miyagawa95}
 slight pressure of 300~bar
is sufficient to reach the superconducting state with $T_c\approx
12.8$~K.\cite{Williams90} For the Mott
insulator \etcn\ hydrostatic pressure of 1.5~kbar is required to
enter the superconducting state at 2.8~K.\cite{Geiser91}

\etcn\ triggers particular interest because at ambient pressure no
indication of magnetic order could be observed down to lowest
temperatures, despite the considerable antiferromagnetic exchange
of $J\approx 250$~K within the triangular lattice;
\cite{Shimizu03,Kurosaki05} thus it is considered as the first
realization of a quantum spin-liquid state suggested by Anderson
40 years ago.\cite{Anderson73} Numerous theoretical and
experimental work was performed during the last decade in order to
explore mainly the thermodynamic and magnetic
properties.\cite{Yamashita08a,Yamashita08b,Pratt11,Nakamura09,Kandpal09,Balents10}
In the high-temperature range, the $^1$H-NMR relaxation rate shows
anomalies around 200~K,\cite{Kurosaki05} the thermopower at
150~K,\cite{Komatsu96} microwave experiments exhibit a dielectric
anomaly at 113~K,\cite{Poirier12} while a peak in
$\epsilon^{\prime}(T)$ occurs below 60~K in the radio-frequency
range.\cite{AbdelJawad10} In addition, a low-temperature anomaly
near 6~K has been observed in
thermodynamic,\cite{Yamashita08a,Shimizu03}
transport,\cite{Yamashita08b} dielectric \cite{Poirier12} and
lattice \cite{Manna10} properties that has not been satisfactorily
explained.

As far as the charge
dynamics is concerned, Ng and Lee suggested that the gapless
spinons contribute to the optical conductivity inside the charge
gap.\cite{Ng07} In the low-frequency and low-temperature limit,
we do in fact observe power-laws, that cross over from a linear to a quadratic dependence.
Unexpectedly and even more striking, however, is the experimental evidence of a
power-law behavior in the frequency dependent conductivity of \etcn\
that extends all the way up to the mid-infrared. The exponent $n$ in
$\sigma(\omega)\propto \omega^n$, slightly rises from with $n\approx 1$ to 1.5
as the temperature is reduced and then saturates for $T<50$~K.

\section{Experimental Details}
Single crystals of \etcn\ (abbreviated as $\kappa$-CN) and \etcl\
($\kappa$-Cl  hereafter) were grown by
electrochemical methods 
and can reach up to
$2\times 2~{\rm mm}^2$ in size with nicely shining surfaces.
Temperature dependent optical reflection experiments $R(\nu,T)$
were performed in a wide frequency range using several Fourier
transform infrared spectrometers ($23-12\,000$~\cm) with an
infrared microscope attached to it;\cite{Dressel04,Faltermeier07}
in addition the high-frequency optical properties (up to
35\,000~\cm) were determined by spectroscopic ellipsometry at room
temperature. For $T<100$~K the optical transmission was measured in
the THz range ($18-46$~\cm) using a coherent source spectrometer.\cite{Gorshunov05} In all cases, the light was polarized in the two main
directions of the crystal planes.
In order to perform a Kramers-Kronig analysis,\cite{DresselGruner02}
the data were extrapolated at low-frequency according to the dc
conductivity measured by the standard four-probe method.
Alternative approaches by $R(\nu\rightarrow 0)={\rm const.}$ and
Hagen-Rubens extrapolation (for elevated temperatures) were tested, but
yield only minor
changes for $\nu<30~$\cm\
and basically no modifications above.

\begin{figure}
\centering
 \includegraphics[width=1\columnwidth]{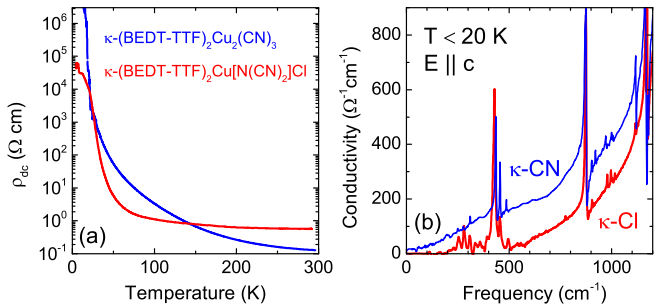}
\caption{\label{fig:2} (Color online) (a) The temperature
dependence of the in-plane dc resistivity  of \etcn\ and \etcl\
evidences an insulating behavior. (b) The low-temperature optical
conductivity ($\kappa$-CN: $T=13$~K; $\kappa$-Cl:  $T=20$~K) does
not show a clear-cut Mott gap, but also reveals important
differences between both compounds.}
\end{figure}
The temperature dependent resistivity of $\kappa$-CN and
$\kappa$-Cl plotted in Fig.~\ref{fig:2}(a) demonstrates the
insulating behavior of both compounds, but also important
differences: cooling from room temperature $\kappa$-Cl crosses
over from a semiconducting phase to a Mott insulator at
$T_M\approx 40$~K and orders antiferromagnetically around $T_N=25$~K.\cite{Yasin11}
For $\kappa$-CN the resistivity continuously
increases upon cooling without any anomaly. No single activation
energy can be extract;\cite{Komatsu96,Kezsmarki06} but  below
approximately 100~K the slope of the Arrhenius plot
$\rho(T)\propto\exp\{\Delta/T\}$ corresponds to $\Delta=200$~K,
right in the range of the exchange coupling $J$. Although the
in-plane optical conductivity of both compounds
[Fig.~\ref{fig:2}(b)] vanishes for low frequencies and
temperatures, the in-gap conductivity is significantly larger for
$\kappa$-CN compared to $\kappa$-Cl. When extrapolating
$\sigma_1(\omega)$ linearly to zero, we may identify the Mott gap around
500~\cm\ for $\kappa$-Cl at
$T=20$~K; however, it is not as well pronounced and with stronger
vibronic contributions than 
reported form early measurements.\cite{Kornelsen92}

\section{Results and Analysis}
\begin{figure}
\centering
 \includegraphics[width=1.0\columnwidth]{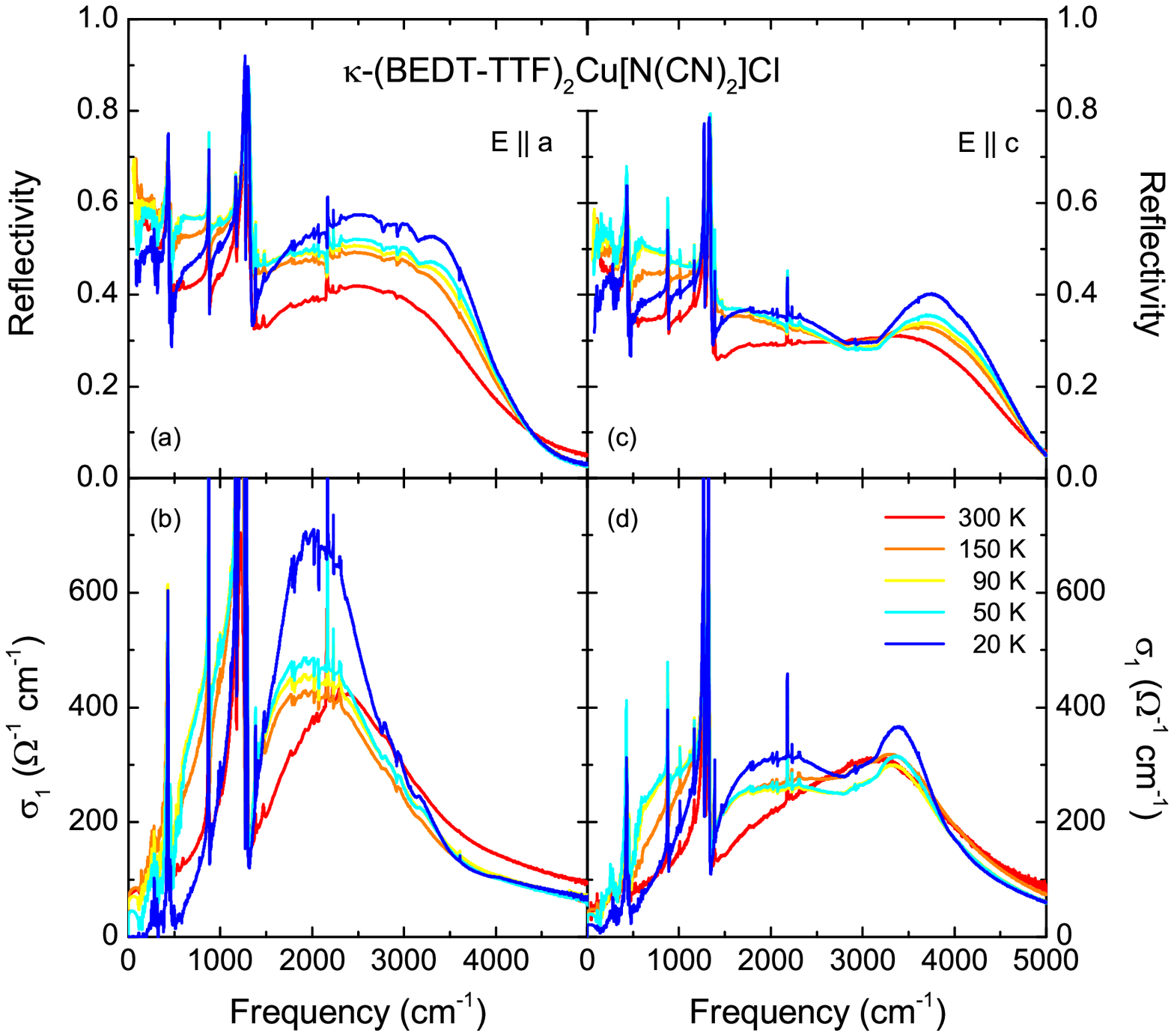}
\caption{\label{fig:S2} (Color online) Evolvement of the optical reflectivity and
conductivity of \etcl\ as the temperatures is varied. For both
polarizations of the light, a gap opens upon cooling and the spectral weight shifts to
higher energies. The data taken from Ref.~\onlinecite{Faltermeier07}.}
\end{figure}
In Fig.~\ref{fig:S2} the frequency-dependent reflectivity and conductivity of $\kappa$-Cl is plotted for different temperatures and polarizations. Since the optical properties of
$\kappa$-(BE\-DT\--TTF)$_2$\-Cu\-[N\-(CN)$_{2}$]\-Br$_{x}$Cl$_{1-x}$
($0\leq x<1$) have been  presented and extensively analyzed during the last years, we refer to Refs.~\onlinecite{Faltermeier07,Merino08,Dumm09,Dressel09} for a detailed discussion.

Here we
concentrate on the spin-liquid compound.
\begin{figure}
\centering
\includegraphics[width=1.0\columnwidth]{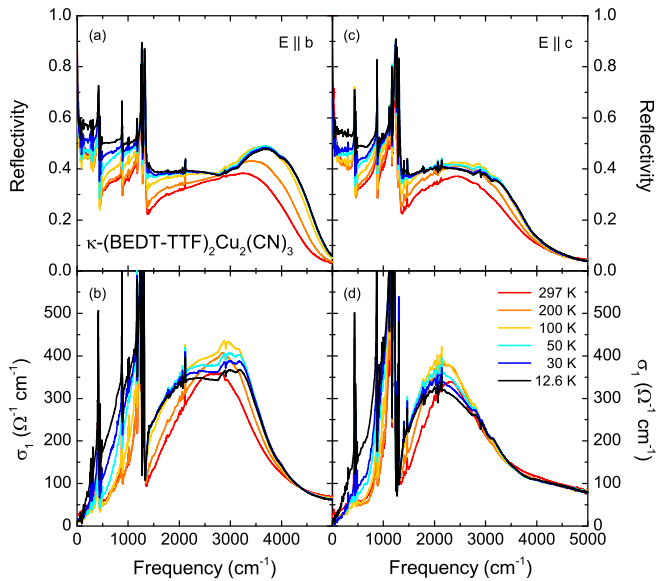}
\caption{\label{fig:3} (Color online) Optical reflectivity and
conductivity of \etcn\ measured at different temperatures along
the two in-plane polarizations as indicated. As the temperature is lowered the low-frequency
reflectivity and conductivity rises, very much in contrast to  \etcl\ displayed in Fig.~\ref{fig:2}.}
\end{figure}
In Fig.~\ref{fig:3} the infrared reflectivity and conductivity of
$\kappa$-CN are displayed for different temperatures; the data are
in accord with previous measurements.\cite{Kezsmarki06} The
optical response is governed by a broad peak in the mid-infrared
caused by transitions between the lower and upper Hubbard band and
intra-dimer excitations; in addition a large number of vibrational
modes extend down to 100~\cm.  The transition between the Hubbard
bands peaks around 2100~\cm\ while the intra-dimer excitations
center around 3000~\cm\ (more pronounced for $E\parallel b$); this
is marginally lower in energy than in
$\kappa$-Cl.\cite{Faltermeier07} Calculations by density
functional theory and tight binding methods yield an intra-dimer
transfer integral $t_1\approx170$~meV with a slight increase upon
cooling,\cite{Jeschke12} in accord with our observations. The
Hubbard band, on the other hand, shifts to lower energies as the
temperature is reduced.

\subsection{Mott gap}
\begin{figure}
\centering
 \includegraphics[width=1\columnwidth]{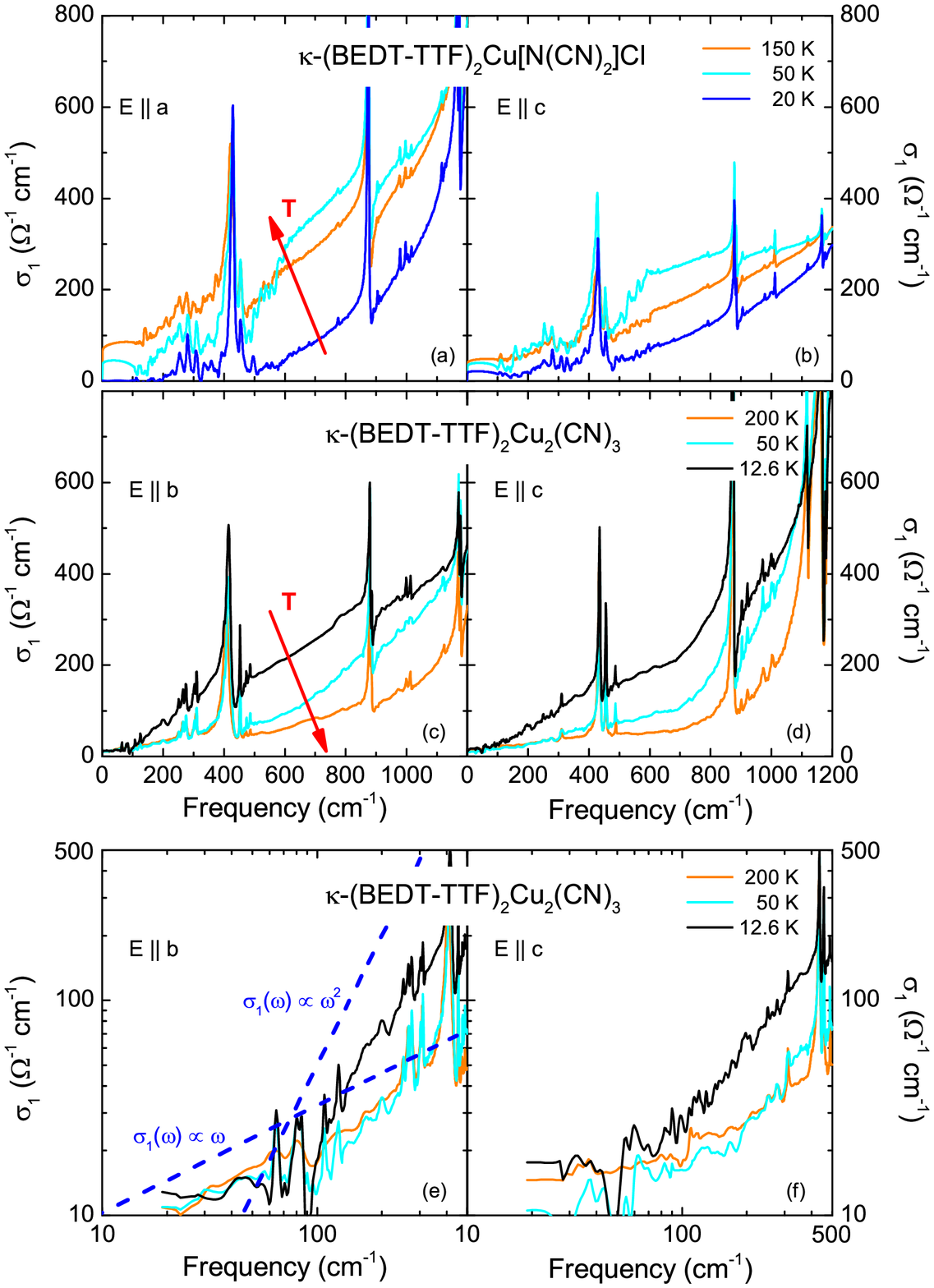}\vspace*{3mm}
\caption{\label{fig:S3} (Color online) Optical conductivity of
(a,b) \etcl\ and (c,d) \etcn\ for  the two in-plane polarizations.
The temperature evolution is completely different for the two
compounds, as indicated by the red arrows. (e,f) The low-frequency
conductivity of \etcn\ plotted in a double logarithmic fashion in
order to demonstrate the power-law behavior. The blue dashed lines
correspond to $\sigma_1(\omega)\propto \omega$ and
$\sigma_1(\omega)\propto \omega^2$, respectively.}
\end{figure}
While at first glance the optical spectra of both Mott-insulators
might look the same at room temperature, there are significant
differences as far as the temperature dependence is concerned;
Fig.~\ref{fig:S3} makes it particularly clear. In the case of
$\kappa$-Cl the Mott insulating state gradually develops upon
cooling, seen in Fig.~\ref{fig:S3}(a,b) by the drop of the optical
conductivity in the far-infrared range. Accordingly the spectral
weight shifts to higher frequencies as the temperature is
reduced.\cite{Dumm09} The Mott gap increases upon cooling and
arrives at $\Delta_{\rho}(k_B/h c)\approx 500$~\cm\ for $T=20$~K
and might become slightly larger in the $T\rightarrow 0$ limit.
Systematic investigations as a function of effective correlations
$U/t$ by applying chemical
pressure\cite{Faltermeier07,Merino08,Dumm09} confirm this
assignment.

The optical conductivity of $\kappa$-CN, however, exhibits the
{\em opposite temperature dependence} [Fig.~\ref{fig:S3}(c,d)],
and it is not possible to identify a clear-cut energy gap from our
optical measurements. While in the low-frequency limit
($\nu\rightarrow 0$) the conductivity decreases upon cooling
[according to the insulating behavior observed in the dc
resistivity, Fig.~\ref{fig:2}(a)], above approximately 100~\cm\
the infrared conductivity  actually rises. This is in stark
contrast to the opening of an energy gap inferred from $\rho(T)$;
instead it indicates that additional excitations develop for low
temperatures, which extend down to small frequencies. In cuprates,
a pseudogap develops upon cooling even in the superconducting
systems.\cite{Timusk99} In the present case of the quantum
spin-liquid compound $\kappa$-CN, however, the conductivity
behaves completely different with a strong frequency and
temperature dependence of $\sigma_1(\omega,T)$.

\subsection{Spinon excitations}
In order to explain the strong in-gap excitations,
Ng and Lee suggested that due to coupling with the internal
gauge field, spinons may contribute to the optical conductivity of
a spin liquid.\cite{Ng07}
They predict a strongly enhanced conductivity
within the Mott gap and a power-law absorption at low frequencies, i.e.\
for energies smaller than the exchange coupling $J\approx 250$~K.
For low frequencies ($\hbar\omega < k_BT$) the
optical conductivity $\sigma_1(\omega)\propto\omega^2$, while for
$\hbar\omega
> k_BT$ the power law should increase to $\sigma_1(\omega) \propto
\omega^{3.33}$.

In order to check whether such a behavior can indeed be found in the spin-liquid compound
$\kappa$-CN, we first tried to fit our conductivity data directly
by the expected power law $\sigma_1(\omega) \propto
\omega^{n}$. The double-logarithmic plot in Fig.~\ref{fig:S3}(e,f) allows us
to readily determine the exponent $n$: for high temperatures we obtain $n\approx 1$ at small frequencies
with a slight increase to $n\approx 1.5$ above 150~\cm. For $T=12.5$~K, $\sigma_1(\omega)$ rises even slower than linear at low frequencies with a crossover to $n=2$ for $\nu>70$~\cm.
The power-law exponents are slightly smaller for $E\parallel c$ compared to $E\parallel b$.
While we find a qualitative agreement with theory of optical excitation of  gapless spinons, the experimentally obtained power laws are weaker by a factor of two compared to the proposed ones.

\subsection{Disentangling the spectra}
As can be seen in Figs.~\ref{fig:2} and \ref{fig:S2},
the spectra of $\kappa$-CN and $\kappa$-Cl are rich in vibrational excitations. Most of them are due to internal vibrations of the BEDT-TTF molecules, with many of them excited via electron-molecular vibrational (emv) coupling \cite{Dressel04} which results in asymmetric Fano  resonances. They are best taken into account
by the cluster model (see for example Rice\cite{Rice76} and Delhaes and
Yartsev \cite{Yartsev93}) that describes the optical properties of
molecular clusters with arbitrary geometry and equilibrium charge
density distribution.\cite{Faltermeier07,Vlasova09} In addition, vibrations of the anion and lattice vibrations are present.

Due to   strong electronic correlations, in these Mott insulators
itinerant carriers are absent  at low temperatures, as discussed in detail in Ref.~\onlinecite{Faltermeier07}.
Thus except for elevated temperatures, no Drude-like contribution is found in the spectra.
Transitions between the Hubbard bands can be identified in the optical conductivity
as maxima slightly above 2000~\cm.
Intradimer transitions cause a band around 3000~\cm\ that shows up as a shoulder or separate maximum. Both contributions form the dominant mid-infrared band;
we describe these excitations by the Lorentzian model.

\begin{figure}
\centering
 \includegraphics[width=0.8\columnwidth]{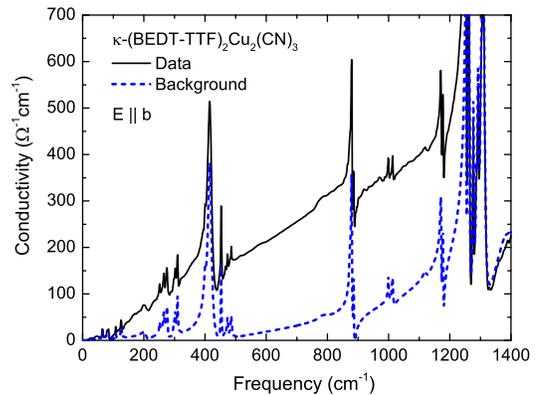}
\caption{\label{fig:S4} (Color online) Optical conductivity of
\etcn\ ($E\parallel b$, $T=12.6$~K) and background due to
molecular and lattice vibrations and interband transitions, which
is subtracted in order to retrieve the possible electronic and spinon
contributions.}
\end{figure}

The uncoupled vibrational features and electronic excitations were fitted by the Lorentz model \cite{DresselGruner02}
\begin{equation}
\hat{\sigma}^{\rm Lorentz} =\sigma_1(\omega) +{\rm i}\sigma_2(\omega)  =\frac{N e^2}{m}  \frac{\omega}{{\rm i}(\omega_0^2-\omega^2)+\gamma\omega}
\label{eq:Lorentz}
\end{equation}
centered around the eigenfrequency $\omega_0$ with a width $\gamma=1/\tau$.
Depending on the interaction of the vibrational excitations with the electronic background,
the modes may become asymmetric and have to be modeled by \cite{Fano61,Damascelli97}
\begin{equation}
\hat{\sigma}^{\rm Fano} = {\rm i}\sigma_0(q-{\rm i})^2({\rm i}+x)^{-1}
\quad ,
\label{eq:Fano}
\end{equation}
where $\sigma_0$ is the background,
$x=(\omega^2-{\omega_0}^2)/\gamma\omega$ ($\gamma$ and $\omega_0$
are the linewidth and the resonant frequency, respectively) and
$q$ is the Fano parameter reflecting the degree of asymmetry of
the peak. As an example, in Fig.~\ref{fig:S4} the optical
conductivity for $\kappa$-CN is displayed as measured at
$T=12.6$~K for $E\parallel b$; we also plot the background
contributions due to intra- and inter-molecular vibrational
excitations, due to transitions between the Hubbard bands and
intra-dimer excitations.

In order to disentangle the various contributions we simultaneously fitted the measured reflectivity and conductivity spectra in the frequency range below $1100-1400$~\cm\ (depending on temperature and polarization) by
\begin{eqnarray}
\hat{\sigma}(\omega) &=& \hat{\sigma}^{\rm background}(\omega) + \hat{\sigma}^{\rm res}(\omega) \nonumber \\
&=&  \sum_{i} \hat{\sigma}_i^{\rm Lorentz} + \sum_j \hat{\sigma}_j^{\rm Fano} +
\hat{\sigma}^{\rm res}(\omega)  \quad ,
\label{eq:fit}
\end{eqnarray}
where we used a minimum number of Lorentz terms $\hat{\sigma}_i^{\rm Lorentz}$ and of Fano terms $\hat{\sigma}_i^{\rm Fano}$.  For the residual conductivity contribution we added two power-law terms with different prefactors $A$ and $B$ and exponents $n$ and $m$:
\begin{equation}
\sigma_1^{\rm res} = A\, \omega^n |_{\rm \omega<\omega_c} +B\, \omega^m |_{\rm \omega>\omega_c} \quad .
\end{equation}
The crossover frequency between the regions of two power-law exponents was found to be at 600~\cm\ at $T=300$~K and increasing to 800~\cm\ for low temperatures. The automated procedure was based on the root-mean-square deviation of the fit compared to the experimental data. The uncertainty for the exponent was estimated to be $\pm 0.2$ and indicated by error bars in Fig.~\ref{fig:8}.

\subsection{Power laws}
\begin{figure}
\centering
 \includegraphics[width=\columnwidth]{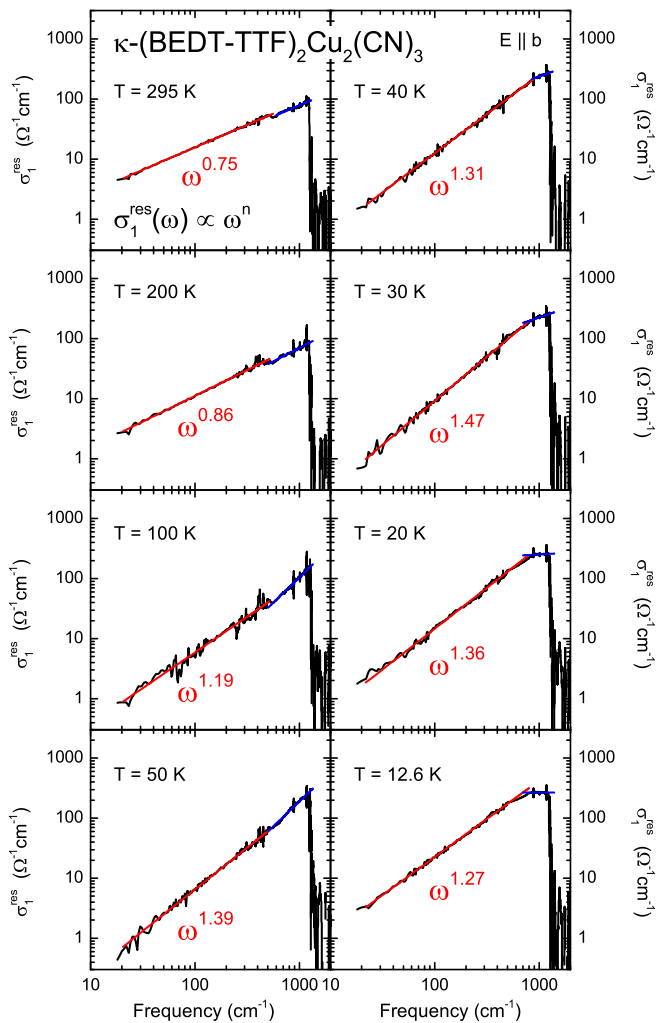}
\caption{\label{fig:7} (Color online) Frequency dependent
conductivity of \etcn\ after the background due to vibrational and
interband excitations has been subtracted shown. The experiments
were performed for the polarization $E\parallel b$ at different
temperatures as indicated. A power-law behavior $\sigma_1(\omega)-\sigma_1^{\rm background}(\omega)= \sigma_1^{\rm
res}(\omega) \propto \omega^n$ can be extracted over a large frequency
range (indicated by red lines). A change in slope may be
identified for higher frequencies (blue lines) that becomes in
particular obvious at low temperatures. }
\end{figure}
The residual conductivity
$\sigma_1^{\rm res}=\sigma_1-\sigma_1^{\rm background}$ is plotted in
Fig.~\ref{fig:7} for several temperatures. The same analysis can
be done for the polarization $E\parallel c$, leading to very
similar results. Above 20~\cm\ the  conductivity  follows a
power-law $\omega^n$ over a very large range which extends to
600~\cm\ at room temperature and up to even 800~\cm\ at low
temperatures. Above that frequency, a different exponent may be
identified in a limited range up to 1200~\cm. Most important,
there is definitely no increase in slope observed when going from
low to high frequencies; we cannot identify any crossover to occur
at a frequency $\hbar\omega \approx k_B T$, i.e.\ below 200~\cm.

\begin{figure}
\centering
 \includegraphics[width=1\columnwidth]{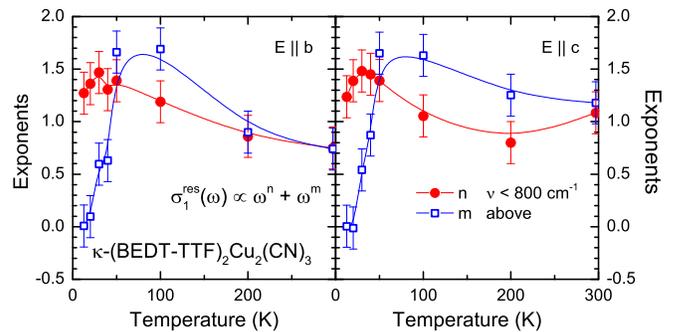}
 \caption{\label{fig:8} (Color online) Exponents $n$ (red dots) and $m$ (blue squares)
extracted from the power law $\sigma_1^{\rm res} = A\, \omega^n
|_{\rm \omega<\omega_c} +B\, \omega^m |_{\rm \omega>\omega_c}$
measured at different temperatures for the polarizations
$E\parallel b$ and $E\parallel c$. The lines correspond to spline
interpolations. Ng and Lee (Ref.~\onlinecite{Ng07}) predict a
quadratic behavior at low frequencies that crosses over to an even
stronger dependence of $n = 3.33$ at high frequencies due to
spinon contributions to the conductivity. }
\end{figure}

In Fig.~\ref{fig:8} we summarize the temperature dependence of the exponent $n$ obtained in the high and low-frequency range for the two directions of polarization. It starts with a linear dependence $\sigma_1^{\rm res}\propto \omega$ at ambient temperature, but then increases to almost $n\approx 1.5$ when approaching $T\approx 50$~K. A noticeable change is observed when cooled down further: the power-law exponent saturates or even decreases to about 1.25. Taking the uncertainty of determination into account, the low-temperature value is approximately $n=1.35\pm0.1$ for both directions. Interestingly, also the higher-frequency exponent significantly drops for $T<50$~K, and $\sigma_1^{\rm res}(\omega)\approx {\rm const.}$ below 20~K. At no time the power-law in conductivity approaches a quadratic behavior in frequency or even larger exponents $n$ or $m$.

The situation is completely different for the Mott-insulator
$\kappa$-Cl. From Fig.~\ref{fig:2}(b) it becomes obvious that the
optical conductivity is substantially smaller in the spectral
range $\nu<1000$~\cm. The Mott gap is well defined for both
polarizations and it continuously increases   to 500~\cm\ as the
temperature decreases [Fig.~\ref{fig:S3}(a,b)]. The up-shift in spectral weight and the reduced
contribution of the conduction electrons below 50~K is rather a
consequence of the Mott state than magnetic order.\cite{Yasin11,Dumm09,remark1}
Tuning the effective correlations  $U/t$ by chemical
\cite{Faltermeier07,Dumm09} and hydrostatic pressure \cite{Drichko06} causes spectral
weight to shift to lower energies and fill the gap. Although the
room-temperature conductivity appears similar for both compounds, the further evolvement is distinct: for $\kappa$-Cl it is not possible to extract a power-law behavior $\sigma_1^{\rm
res}(\omega)\propto \omega^n$ over a sufficiently large frequency
range for any temperature $T<300$~K.

\section{Discussion and Conclusion}
The low-frequency behavior highlighted in Fig.~\ref{fig:S3}(e,f) could in fact be
caused by optical excitations of gapless spinons, as suggest by Ng and Lee,\cite{Ng07}
although the conductivity should increase much faster and exhibit a stronger frequency dependence than it actually does:
the low-temperature exponents extracted for the power law are significantly lower than predicted. At $T=12.6$~K there is a crossover around 70~\cm\ where the linear rise increases to a quadratic behavior in frequency.
When taking out the vibrational features and interband transitions, this behavior continues all the way up to the midinfrared spectral range: this casts some doubt to this explanation. Even more surprising is the fact, that we do observe the power-law behavior up to room temperature where light-induced spinon excitations should not be observable.
A final decision of the applicability of the model can only be based on advanced experiments at lower frequencies and lower temperatures.


The dependence $\sigma_1^{\rm res}(\omega,T) = A(T)\, \omega^n$
observed for the conductivity of quantum spin-liquid compound
$\kappa$-CN could in fact be caused by fluctuations. The
contribution $A(T)$ significantly increases with lowering
temperature which infers quantum fluctuations. In the vicinity of
a quantum critical point,\cite{Belitz05,Lohneysen07} power-law behavior is
often observed and quantum fluctuations become more pronounced
when $T$ is reduced. We have to keep in mind, however, that this behavior is observed
up to rather high frequencies and at elevated
temperatures, far beyond the regime quantum fluctuations should be
dominant.

It is also not clear why this behavior
saturates around 50~K, since no magnetic, charge, or structural
order is revealed at any
temperature. It coincides with anomalies in the dielectric
properties.\cite{Poirier12,AbdelJawad10} Hotta proposed a
dipolar-spin liquid \cite{Hotta10} assuming quantum electric
dipoles on the dimers that interact with each other through the
dipolar spin coupling. This would lead to a ferroelectric-like
behavior where a soft mode moves to low frequency. However, there
are no indications of charge order in these $\kappa$-phase
compounds,\cite{Tomic12,Sedlmeier12} and soft modes and collective charge
excitations commonly show up at much lower frequencies.\cite{LinesGlass77,Dressel10}
We suggest that charge fluctuations are strongly enhanced in the vicinity of a quantum critical point and correspond to a much larger energy scale than commonly observed in the case of quantum fluctuations in the spin degree of freedom.

Most important, a very similar dielectric anomaly was observed in the in-plane and out-of-plane properties of $\kappa$-Cl\cite{Pinteric99,Lunkenheimer12,Tomic12} where
the Mott gap is clearly developed at low temperatures and no in-gap contribution  are present in the optical conductivity.
Tomi{\'c} and collaborator suggested that grain boundaries, linked to magnetically ordered domains in $\kappa$-Cl might move, thus causing a strong dielectric contribution.\cite{Tomic12} This argument, however, does not hold for $\kappa$-CN where no magnetic order is present down to lowest temperatures. Pressure dependent investigations on $\kappa$-Cl  showed that the bandwidth-controlled Mott criticality involves critical fluctuations in charge,\cite{Kagawa05} spin,\cite{Kagawa09} and lattice.\cite{Fournier03,deSouza07}
Nothing like that is known for the spin-liquid compound $\kappa$-CN for the intermediate temperature range.
While this issue can be clarified
by further experimental studies, considerable theoretical work is
required in order to describe the power-law behavior the dimer Mott insulator with a quantum spin liquid state, in the low-frequency limit as well as in the far-infrared range.

From our optical investigations of two Mott-insulators with very
similar triangular structure but  different magnetic ground
states, we can conclude that only the magnetically ordered organic
salt \etcl\ exhibits a well defined Mott gap at low temperatures.
The quantum spin-liquid compound \etcn\ exhibits a power-law
behavior in the frequency-dependent conductivity that becomes
stronger as the temperature decreases. We suggest spin
fluctuations get important in the spin-liquid state and couple to
the electrodynamic properties differently compared to an
antiferromagnetic Mott insulator. The power-law $\sigma_1(\omega)$
observed in an extended range of frequency and temperature
remains a puzzle that calls for further investigations
of how charge fluctuations are influenced by the presence
or absence of magnetic order.

\acknowledgements We thank R. Beyer for many discussions, P.A. Lee
and R.H. McKenzie for helpful comments. D.W. acknowledges support
by the Alexander von Humboldt foundation. The project was
supported by the Deutsche Forschungsgemeinschaft (DFG). Work
supported by UChicago Argonne, LLC, Operator of Argonne National
Laboratory ("Argonne"). Argonne, a U.S. Department of Energy
Office of Science laboratory, is operated under Contract No.
DE-AC02-06CH11357.

\end{document}